\documentclass[utf8]{FrontiersinHarvard} % for articles in journals using the Harvard Referencing Style (Author-Date), for Frontiers Reference Styles by Journal: https://zendesk.frontiersin.org/hc/en-us/articles/360017860337-Frontiers-Reference-Styles-by-Journal
\usepackage{url,hyperref,lineno,microtype,subcaption}
\usepackage[onehalfspacing]{setspace}

\def\keyFont{\fontsize{8}{11}\helveticabold }
\def\firstAuthorLast{Balasubramanian {et~al.}} %use et al only if is more than 1 author
\def\Authors{Rama Balasubramanian\,$^{1,3}$, Danielle Findley-Van Nostrand\,$^{2}$ and Matthew C.\ Fleenor\,$^{1,4*}$}
\def\Address{$^{1}$Department of Mathematics, Computer Science, \&\ Physics, Roanoke College, Salem, VA, USA \\
$^{2}$Department of Psychology, Roanoke College, Salem, VA, USA \\
$^{3}$ Chance to Change Lives (CCL-US), Pittsburg, PA, USA\\
$^{4}$Department of Chemistry \& Physics, Univ.\ of Mary Washington, Fredericksburg, VA, USA }
\def\corrAuthor{Matthew C.\ Fleenor}
\def\corrEmail{mfleenor@umw.edu}

\begin{document}
\onecolumn
\firstpage{1}

\title[Programmatic Innovations]{Programmatic Innovations that Accord with the Retention of Women in STEM Careers} 

\author[\firstAuthorLast ]{\Authors} %This field will be automatically populated
\address{} %This field will be automatically populated
\correspondance{} %This field will be automatically populated

\extraAuth{}% If there are more than 1 corresponding author, comment this line and uncomment the next one.
%\extraAuth{corresponding Author2 \\ Laboratory X2, Institute X2, Department X2, Organization X2, Street X2, City X2 , State XX2 (only USA, Canada and Australia), Zip Code2, X2 Country X2, email2@uni2.edu}

\maketitle

\begin{abstract}
Gender representation in the physical sciences remains inequitable and continues to lag behind other fields.  Even though there exists adequate documentation regarding programmatic postures, difficulties persist within the physics discipline.  In this paper, we present innovative, programmatic elements over an eight-year period at an undergraduate, liberal arts, physics program.  These elements were added in response to the following two questions: ``What practices cultivate an increase of physics major numbers in an undergraduate, liberal arts setting?''; and, ``What practices facilitate a depth of experience for individual physics graduates?'' Some of these innovations accord with published, `best practices' for undergraduate physics programs, while others are novel to the program's context.  Within this eight-year period, alterations are separated into curricular and co-curricular elements. Innovations are introduced in some detail, and data are presented before, during, and after their introduction.  While it is currently impossible to say which elements had the greatest impact, the synergistic combination did have a positive effect on the program.  Not only did the number of total majors and graduates increase, there was a 200\% increase of women degree recipients compared to the previous ten years, which boosted average graduation rates above the national average ($30 \% > 20\%$).  Women were retained within the undergraduate physics major at a higher percentage during this time period when compared to men in the program.  Lastly, these women physics majors maintained careers in science advancement fields at a rate of $80+$\% after $\leq5$ years post-graduation.  While this paper presents a singular case study, the purpose is two-fold: a) to validate quantitatively the work of national physics organizations within the context of a liberal arts institution, and b) to suggest that a multi-level approach is most efficacious when considering programmatic innovations.
%%% Leave the Abstract empty if your article does not require one, please see the Summary Table for full details.

\tiny
 \keyFont{ \section{Keywords:} physics education, undergraduate experience, women in physics, retention, STEM identity, STEM belonging} %All article types: you may provide up to 8 keywords; at least 5 are mandatory.
\end{abstract}

\section{Introduction}
\subsection{Global Landscape}
One of seventeen United Nations Sustainable Development Goals reads, ``achieve gender equality and empower all women and girls'' \citep{ung23}. Within purportedly more rational and logical disciplines in Science, Technology, Engineering, and Mathematics (STEM), one might expect a better fulfillment of this goal. Yet, STEM disciplines reveal a glaring ``gender gap'' at the professional level, despite the increase of women undertaking higher-education STEM studies. Gender gaps express themselves as differences ``between women and men in terms of their levels of participation, access, rights, remuneration or benefits'' \citep{wef20}. Education is one of the four key areas of identification and measurement. Within the sciences, fewer than 30\% of the world's researchers are women, which reflects a clear gender gap \citep{une19}. Through a conglomeration of global scientific organizations, individual survey results from global scientists, and publication pattern information, a deep synthesis by \citet{roy20} reveals a persistent and significant gender gap within the sciences globally. Sampling equally over 30,000 men and women from 159 countries, this gap remains regardless of STEM discipline, geographic location, or economic development. While what follows is a focused examination of one physics program at an undergraduate institution in the United States (US), the local environment of the study and the national context of the institution are representative of the global landscape.

\subsection{National Landscape }\label{natlScape}

Over the past thirty years in the US, gender representation in the physical sciences continues to be disproportionately skewed against women \citep{ivi12}, despite a three-fold increase in the percentage of undergraduate physics degrees awarded to women (6\% in 1970 versus 22\% in 2018, \citealp{por19}).  According to a 2021 report by the American Physical Society (APS) on building America’s workforce, the importance of developing an inclusive and diverse workforce is crucial to boost the innovation and productivity of science and technology and to maintain America as a global leader in these areas \citep{joh21}. Problems persist in gender equity issues within most STEM fields (e.g., \citealp{hil10}), though physics remains one of the most inequitable.  At approximately 20\% women recipients of undergraduate degrees, physics continues to lag behind mathematics (40\%), chemistry ($\sim$50\%), and biology (60\%).  This considerable difference is displayed clearly in \citet{por19}, where the percentage of women physics-degree recipients remains approximately constant since 2000. Within this same climate, the number of undergraduate-degree recipients is more balanced.  For example, the National Center for Education Statistics reports that women receive 58\% of all undergraduate degrees, while only 36\% of those are awarded to STEM majors \citep{nces19}.

There are various potential reasons for this continued gender disparity in the physical sciences, including stereotypes relating the practice of physical science to hyper-masculinity \citep{fra17,goo08,smy15}, or a lack of perceived representation of women or women role models in the field, in turn leading to issues of belonging and fit \citep{nel10,gis18,goo12}.  There are also individual-level factors to consider, including women’s lower self-efficacy for physics (not coupled with lower objective skills; \citealp{kal20}), their experienced science identity \citep{ere21,tru14}, or a perceived lack of opportunity to fulfill communal career goals, which tend to be valued more by women \citep{die17}. 

Simultaneous to persistent gender inequity for physics undergraduate recipients, US organizations like the Statistical Research Center of the American Institute of Physics (AIP) address and publicize these known inequities through detailed study, survey, and site visitation \citep{src21}.  The AIP continues to produce valuable statistics on historically underrepresented populations in physics specifically, and STEM fields as a whole.  This important work presents the results of labor within the field, as well as the persistent gaps and needed future focus.  In addition, commissions of specialists within the discipline have produced and published manuals containing guidelines on undergraduate program efficacy.  This is particularly true for undergraduate programs as a whole (SPIN-UP, \citealp{spin03}), career preparation (J-TUPP, \citealp{jtup16}), and African-American undergraduates within physics (TEAM-UP, \citealp{team20}).  Resources related to programmatic guidance, change, and growth are available through the interactive AAPT-EP3 website, supported by the American Association of Physics Teachers \citep{ep321}.  It is essential that physics programs consult and digest the results of these resources and apply the general guidelines to their specific institutional contexts.

Because the current study focuses on an undergraduate physics program in the US, some differences between traditional American and non-US undergraduate universities are made clear.  Most popularly, it is well-known that an undergraduate degree in the US takes approximately one-year longer than systems outside of the US, particularly European-based systems.  Part of this prolonging by the US system accords with an emphasis on general education coursework, which usually delays the declaration of an intended undergraduate degree program, usually referred to as a ``major.'' Moreover, student choice of an undergraduate major is a novelty within a US-based, post-secondary education, and it is a source of intense study.  For example, how students choose an undergraduate major is a recent, scholarly focus for economics \citep{wis21}, race and gender studies \citep{rai18}, and social psychology \citep{den21}.  A major declaration is further complicated in a liberal arts environment, which is discussed below in Sec.\ 1.3.  In contrast, choosing a major is not a feature of many/most non-US systems.  Whether in favor of an Dual System, like Germany (e.g., \citealp{nas12}), or a traditional three-year framework, undergraduate degree pursuits outside the US are more streamlined.

Within the context of total undergraduate degrees awarded in the US, physics programs maintain a meager percentage of STEM bachelor's recipients.  Approximately 20\% of all undergraduate degrees are awarded to STEM fields, while approximately 2.2\% of those degrees were awarded to physics over the past 20 years \citep{aps20}.  Even though PhD-granting institutions comprise only one-quarter of all institutions offering a physics bachelor's degree, they award approximately one-half of all physics bachelor's degrees. When these national percentages are applied to the reduced numbers of undergraduate populations at liberal arts colleges, where a STEM culture does not often exist, the corresponding enrollments for STEM courses and physics majors are drastically reduced.

While demand for STEM-degree recipients increases, it may appear to the public that the numbers of physics graduates are insignificant compared to the whole. Even though a small percentage of bachelor's degree recipients will pursue graduate studies in physics, the importance of a physics degree persists through the application of the discipline's transferable skills in other fields \citep{hun13}.  `Hidden physicists' (that is, ``those who are trained in physics but actually work in a job more widely,'' \citealp{jtup16}) populate not only other STEM fields and education landscapes, but also peripheral fields like law and management. Physics degree recipients historically score among the highest on standardized, pre-professional entrance exams like the Medical College and Law School Admission Tests (MCAT and LSAT, \citealp{tes13,tyl22}).  Furthermore, physics degree recipients maintain a privileged position in terms of earning opportunities and employment satisfaction (cf., Fig.\ 3 in \citealp{jtup16}).  For these reasons, along with the continued need for STEM innovation across various career paths, it is vital that physics programs continue to produce degree recipients for careers as `hidden physicists.' This is especially true at institutions where typically less than half of the bachelor's recipients matriculate to graduate physics programs. 

\subsection{Local Landscape}

The current investigation is situated within a context of small number statistics, both nationally within the discipline of physics and locally within an undergraduate, liberal arts college. The physics program discussed here (referred to as the Physics Group) is part of a shared department with mathematics and computer science.  Bachelor's degree-granting, physics programs within shared departments comprise less than 10\% of all US programs \citep{mul21}. Moreover, while offering approximately one-half of the bachelor's degrees, non-PhD granting programs comprise three-quarters of the undergraduate physics landscape. Therefore, a fertile opportunity exists for these programs to affect STEM students with physics content and potential future trajectories.

Beyond the broad differences of matriculation length and focus of study outlined in Sec.\ \ref{natlScape}, a liberal arts setting adds another layer of nuance to the US undergraduate system.  Liberal arts colleges in the US typically provide an undergraduate-focused education emphasizing a general curriculum balanced with humanities and social sciences content, which is required for all students. Both public, state-funded and selective, private-funded institutions are labeled 'liberal arts colleges.'  In addition, most liberal arts institutions have enrollments of less than 5,000 students. Regardless of the differences, the selection of a major is not rushed for undergraduates at these institutions.  It is not uncommon to declare a major in the second year of undergraduate studies at a liberal arts institution, where becoming a (physics) major is a celebrated event.

Many non-PhD granting physics programs contain less than five full-time faculty and produce less than ten graduates per year.  For instance, \citet{tyl20} present statistics demonstrating that a majority of undergraduate-only programs have five or less graduates per year {\em and} five or less full-time equivalent faculty. The Physics Group also finds itself in a similar situation, where it has three tenure-track (TT) members, a non-permanent visiting position (VAP), and a permanent, non-tenure track (NTT) position.  This was also the case ten to fifteen years ago when the Physics Group contained four TT and one NTT faculty, while it awarded $3.6$ degrees per year.  The number of women Physics Group faculty during the time-period of the study was either two (1 TT, 1 NTT), or three (1 TT, 1 NTT, 1 VAP) of five. While the non-tenure track and visiting faculty members primarily taught physics courses for non-majors, they actively contributed to additional learning experiences for women students such as training teaching assistants, troubleshooting lab equipment, facilitating travel to CUWiP conferences.  It is not the focus of the current study to evaluate the effects of faculty gender distribution. The average percentage of degrees awarded to women during the same time period (2003--12) was approximately 20\%, with the national average. When considering published national statistics, the program in this study could be classified as a typical, undergraduate physics program in the US, prior to the study focus (2013--2021). 

The current study maintains the following trajectory: innovative, programmatic elements are introduced and described in the next section.  Next, we provide representation and retention of physics degree recipients before, during, and after the implementation of the programmatic elements.  While causality is not the purpose of the presentation, it will be clear that the increases for the program accord with the implementation period.  A discussion of the impact on the Physics Group in light of these increases follows with some suggestions for other similar programs.  While each institutional context is unique, and the synergistic effect of innovative elements cannot be deduced a priori, the experience of the Physics Group substantiates two demonstrative realities: published guidelines and statistics support programmatic growth goals, and multi-compartment implementation provides an effective programmatic impact.

\section{Materials and Methods}
\subsection{Data Acquisition}
The Office of Institutional Research and the Office of the Registrar were queried for the numbers of majors and graduates for the previous twenty years so that a baseline of physics participation was determined. For the eight years of interest within the study (graduates from 2013--2021), more detailed information was acquired and collated, including course rosters, extra-curricular participation, and post-graduation employment.  This information was accessed through interpersonal communication, social media, and programmatic assessment. The figures within the study were compiled from the results of these data.

The timetable for programmatic alteration was carried out with the following provisos: only one element was implemented in a given year, innovative elements were connected directly to items within the annual programmatic assessment plan, and these elements were carried out for a minimum of three years before conducting evaluation.  This three-year baseline for introduction and implementation of programmatic elements forms a justification for the presentation of data in subsequent sections. Innovative elements were implemented locally based on consultation of the Physics Education Research (PER) literature as well as familiarity with the program, as detailed in the following sections.  The addition of innovative elements for programmatic growth was driven equally by the following two questions: ``What practices cultivate an increase of physics major numbers in an undergraduate, liberal arts setting?''; and, ``What practices facilitate a depth of experience for individual physics graduates?''

Responsibilities for implementing innovative elements were shared among the Physics Group faculty.  Within courses, the instructor of record maintained primary leadership over the specific implementation and data collection. For intra-programmatic or extra-curricular elements, a particular physics faculty member oversaw implementation and collection. While literature-based elements were implemented as accurately as possible, concessions for context and student populations were judiciously applied.  In what follows, physics program alterations are categorized and discussed within the contexts of undergraduate education literature and the particular institutional environment.

\subsection{Introduction of Innovative Elements}\label{innovateIntro}
Programmatic transformation within STEM undergraduate programs is not clear-cut and often dependent on localized factors.  Despite public usage of the STEM acronym, it is clear that even the ``S'' (science) is not monolithic in its practice \citep{mar13}. Moreover, when considering programmatic change initiatives, points of emphasis vary based on discipline \citep{rei19}.  These differences stand outside the unique departmental and institutional cultures on each campus \citep{hen15}.  Detailed case studies of undergraduate programmatic growth exist in the literature for large, US institutions \citep{ste13} and diverse ``thriving'' programs \citep{spin03}.  The `best practices' documentation contains some, but not all, of the innovations that are introduced here. 

Consideration of significant programmatic change began in 2011 while program faculty were preparing for an external review. This physics program review occurred in 2012, during which some of the proposed elements were discussed with evaluators. Accordingly, a focus on growth pivoted from recruitment to retention, because program faculty overlooked that retention could double the number of physics graduates. While some changes were contextual and on-going, major elements were implemented over a four-year period, 2013--2017, along two distinct compartments of the program (curricular and co-curricular).  With a four-year undergraduate matriculation period, eight academic years within the Physics Group provide the basis for the present study (i.e., 2013--21).

\subsubsection{Curricular}\label{curricular}
While faculty may have significant autonomy within the classroom or individual course offerings, instituting change within a curriculum is not an individual escapade.  The process requires input and collaboration from all the physics faculty, where consensus is the aspirational goal.  Moreover, the students themselves must also display a receptivity to any modifications and justifications that are offered.  Here, we present details regarding two major curricular changes.

\textbf{ (A) First-year Colloquium.}  Belonging and science identity are interrelated, drive retention in STEM majors, and may be especially important for women and minority students \citep{goo12,rai18}. Outside of a traditional curriculum, these qualities can be facilitated by affording students early and ample opportunities to connect with one another and with faculty members. The first-year, Physics and Engineering Colloquium meets weekly as an exploratory course emphasizing overarching themes in the physical sciences.  While maintaining a high relational component for cohort-building, the half-credit course is graded on a ``pass-fail'' basis and is based on participation, completion of assignments, and written reflection quality.    Throughout the semester, the first-year students are also introduced to several different cohorts within the physics major (upper-level students and faculty), while also engaging with a breadth of generalized content (order-of-magnitude estimates, physical modeling, and 'how things work').   Hands-on investigations supplement classroom sessions in order to emphasize the experimental aspects of the discipline, as well as increase self-efficacy for tasks needed in subsequent physics courses. In summary, a successful colloquium experience cultivates the following: social capital (cohort-building and inclusion, \citealp{abb05}), content engagement in the discipline (identity), and active learning in the discipline (self-efficacy). By instituting a first-semester course where students of similar interests gather, a like-minded cohort of learners is formed within an inclusive environment, which is supported as a means of establishing a STEM identity and sense of belonging \citep{lew17}.

\textbf{(B) Upper-level Laboratories.}  The Physics Group also made alterations to increase active and applied learning opportunities, as well as facilitate essential experimental skill development. Due to students' interests in applied physics and engineering, two laboratory augmentations were made, while adhering to the college-wide constraints for number of major-only credits.  First, the previous, junior-level, one-semester ``advanced laboratory'' course, which consisted of verifying physical constants, was converted into an intermediate laboratory \citep{alpha}. This laboratory accompanied a third-semester Modern Physics course that introduced the following novel facets:  more developed experimentation and report writing, deepened uncertainty quantification, and emphasized historical and philosophical aspects of science.

With space created at the junior-level, a second laboratory course was added to effectively and efficiently address interdisciplinary topics within the physics major. Following guidance from the literature on Course-based Undergraduate Research Experiences (CUREs; \citealp{cor14,woo18}),  authentic perspective was provided on the specific techniques, while also instructing the students on content that they would not otherwise receive \citep{mor17}.  Students collaboratively completed four mini-research projects in the following areas: Astrophysics, Biophysics, Materials, and Optics/Spectroscopy.  These four areas coincided with four, rotating, upper-level electives. As a result, students were introduced to all interdisciplinary electives through the CURE at a cursory level, even though they will not take all the courses in their entirety.

\subsubsection{Co-curricular}
Guidance from US-based organizations on undergraduate physics education emphasizes several co-curricular strategies to facilitate student and program success \citep{spin03,ep321}. All published studies mention a vibrant Society of Physics Students (SPS) chapter, or similar institution-based student cohort.  With very little previous involvement, the Physics Group became more active with SPS in 2013, though its first year-end Chapter Report was not submitted until 2015. Here, five co-curricular, programmatic innovations are introduced that extend beyond a thriving SPS chapter.

\textbf{(A) Public Science Outreach (Informal Programming).}  Cohort-building is a significant component of the program already mentioned (e.g., the first-year physics colloquium).  Cohort-building implies an individual belonging and inclusion that traditional usage of ``community'' does not \citep{gow14}.  Another way to build interest-based inclusion is through student groups and science outreach to the public \citep{hin16}.  Science outreach opportunities (or `informal programs') not only serve public scientific literacy by raising awareness at an early age, but informal programs also empower undergraduates \citep{ret21}.  When the Physics Group began a concerted informal program effort ten years ago, most of the events were faculty-organized and led.  Within responsible and eager undergraduate leaders, informal programs transformed into a student-led effort.  One such example was the total solar eclipse of 2017, where students served as Eclipse Ambassadors, which resulted in a nationally recognized award (Blake Lilly Prize). After conducting a well-organized event that served over 500 citizens, physics majors presented their experiences to their peers after their return. This further resulted in local news articles about their ambassadorship.  Experiences such as these provide demonstrable opportunities for increased inclusion, efficacy, and identity within physics. 

\textbf{(B) Junior Review.}  A second, related co-curricular addition to the program is Junior Review, an informal interview involving at least two faculty members and the individual physics major.  This addition to the program is beneficial for multiple reasons. First, having multiple faculty members in attendance allows students to participate in collegiality and camaraderie first-hand.  This approach also fosters belonging within an inclusive learning environment, which seems particularly meaningful for women \citep{lew17}.  Secondly, informal questions encourage each student to verbalize the ways and directions in which their interests may have changed (e.g., “In what ways has your interest in physics increased and/or decreased?”). Such self-reflection contributes positively to learning and achievement, and may help develop students’ sense of meaning or purpose within physics \citep{fle18}. Instances of `hidden physicist' trajectories often arose within Junior Review conversations, which encourage new avenues of exploration are not hindered by presumptive assumptions \citep{alo09}. Third, the review is also an opportunity to facilitate participation in ``high-impact practices” tied to deep learning, including research mentored by faculty, supportive minors and/or concentrations, and off-campus internships \citep{jtup16}. 

\textbf{(C) Conference Attendance.}  Prior to 2012, student conference attendance within the Physics group was primarily synchronized with the mentoring faculty researcher.  This was sparse, totaling less than five instances in ten years.  There were many contributing factors, including a lack of faculty attendance, lack of results, and lack of funding.  Beginning in 2012, students attended conferences where undergraduate participation was encouraged regardless of faculty presence (e.g., regional opportunities). The institution developed on-campus poster sessions where students could present their work in less-threatening environments.  With the establishment of a campus-wide Director of Undergraduate Research, monetary funding opportunities for students increased.  These college-wide initiatives led to increased numbers of physics majors attending conferences.

\textbf{(D) Definitions of Physics Excellence.}  Another co-curricular initiative related to the number and definition of year-end physics recognition, for (not-yet) majors.  Traditionally, the institution sponsored one ``Senior Scholar'' award for the highest academic grades. To incorporate a holistic picture of excellence reflecting more than academic achievement, and to facilitate identity and belonging within the discipline, several new awards were added. For example, year-end recognition for majors was given for research within and service to the Physics Group.  First-year awards were given for early achievement in the discipline to those considering a major in physics. These emphases properly reminded students that grades (marks) do not solely determine their undergraduate success, their inclusion within the discipline, nor their future trajectory as a physicist.

\textbf{(E) Experiential Learning.}  Traditional extracurricular research and internship opportunities are widely recognized as best-practices for cultivating STEM identity and belonging, particularly within traditionally underrepresented STEM populations \citep{est18}.  Due to the limited number of research projects within the Physics Group, faculty pursued creative avenues for physics-related extracurricular experiential learning (EEL).  Beyond more common, widely-publicized Research Experiences for Undergraduates (REUs), EEL opportunities for majors within the Physics Group were initiated with regional industry corporations, regional and on-campus collaborators, and Physics Group alumni. An introductory independent study course was also created to better prepare students for their (predominantly) summer EEL participation.  The pre-emptive courses gauged student interest, facilitated research prowess, and built resilience.  By not making a research requirement within the curriculum, the Physics Group invites a student to discover for themselves how best to uniquely experience physics.  These opportunities not only increase the total number of students who participate in EEL, but they also broaden and diversify how a physicist is defined in society.

\section{Results}

Since specific changes and their effects are not isolated, program data are presented before, during, and after implementation.  This precedent follows best-practices by recognizing departmental culture and multiple change agents \citep{dun11,rei19}. Where results pertain to primarily one element, it is recognized that other elements are also `running in the background.'  Entanglement between innovative elements is discussed after the presentation of the increases in programmatic markers.

Two features of the results deserve clarification.  There is a preference to average (or sum) over three-year increments in the data presented.  There are three justifications for this approach.  One, the innovations were staggered and repeated for a three-year timescale before evaluating their effectiveness. Curricular implementations did not initiate in the same academic year, so the three-year average provides an opportunity to see the partial development of one innovation and its integration within the program more holistically. Second, a three-year timescale defines the active trajectory for an undergraduate physics major.  Upper-level core courses are taught on every-other-year basis, which also serves to compress the third and fourth year of the program.  Moreover, by a student's fourth year, much of what they do serves as preparation for post-graduate decisions. While a physics major may undergo some significant intellectual transformation within their fourth year, this is an exception not the rule.  Third, a three-year timescale provides an opportunity to discuss numbers of physics majors (and graduates) in more meaningful quantities. Since the issue of small-number statistics forms the basis for a second clarification, we now transition to that discussion.

Historically, the number of people pursuing physics degrees has always been small in comparison to the whole of undergraduate degrees awarded. Compounded with that fact is the total enrollment at the liberal arts college where the study was conducted ($\sim2000$). Since Poisson distribution statistics characterize samples that are collected at random but with a definite average rate, we believe that they accurately describe the occasion of a woman choosing and matriculating through an undergraduate physics program. The definite average rate is set by the typical four-year matriculation period through the US undergraduate system, where the randomness is based on the gender of a student and their choice of a major. The uncertainties for Poisson samples are provided by $\sqrt{N}$, where $N$ is the number of events. Such uncertainties are valid whether cumulative or average populations are examined, since the standard deviation for a Poisson distribution is also given as $\sqrt{\mu}$, where $\mu$ is the mean count \citep{tay97}. Based on the programmatic justification for three-year increments given above, Poisson uncertainties measure the benefit of innovation implementation above a typical statistical noise.

The use of three-year increments and Poissonian uncertainties are utilized throughout the results of the study.  Therefore, a description is provided for the general presentation of the results section.  In all histograms outlining the numbers of students for time-periods before and after the study, the results are binned according to the same years.  A marker denotes the initiation of the study.  When error bars are provided, they are calculated by the square-root of the number, whether cumulative or average numbers are presented. When other presentations of data are utilized, they are explained in context.

\subsection{Increased Number of Physics Majors \& Graduates}

During the time period of implementation (2012 through 2021), the Physics Group did not have any external changes regarding number of faculty, departmental situation, or physical location.  Over the twenty-year period presented here, there was some faculty turnover though the number of positions remained constant.  With respect to national standards, the Physics Group is considered ``normal'' regarding the number of physics faculty \citep{tyl20}.  How1ever, the number of women faculty could be considered important since it was higher than the national average (40\% compared to 20\%). The Physics Group's building did not change during the period of implementation and remained the oldest academic building without renovation.  

\begin{figure}[!h]
  \begin{center}
\includegraphics[width=17.5cm,height=9cm]{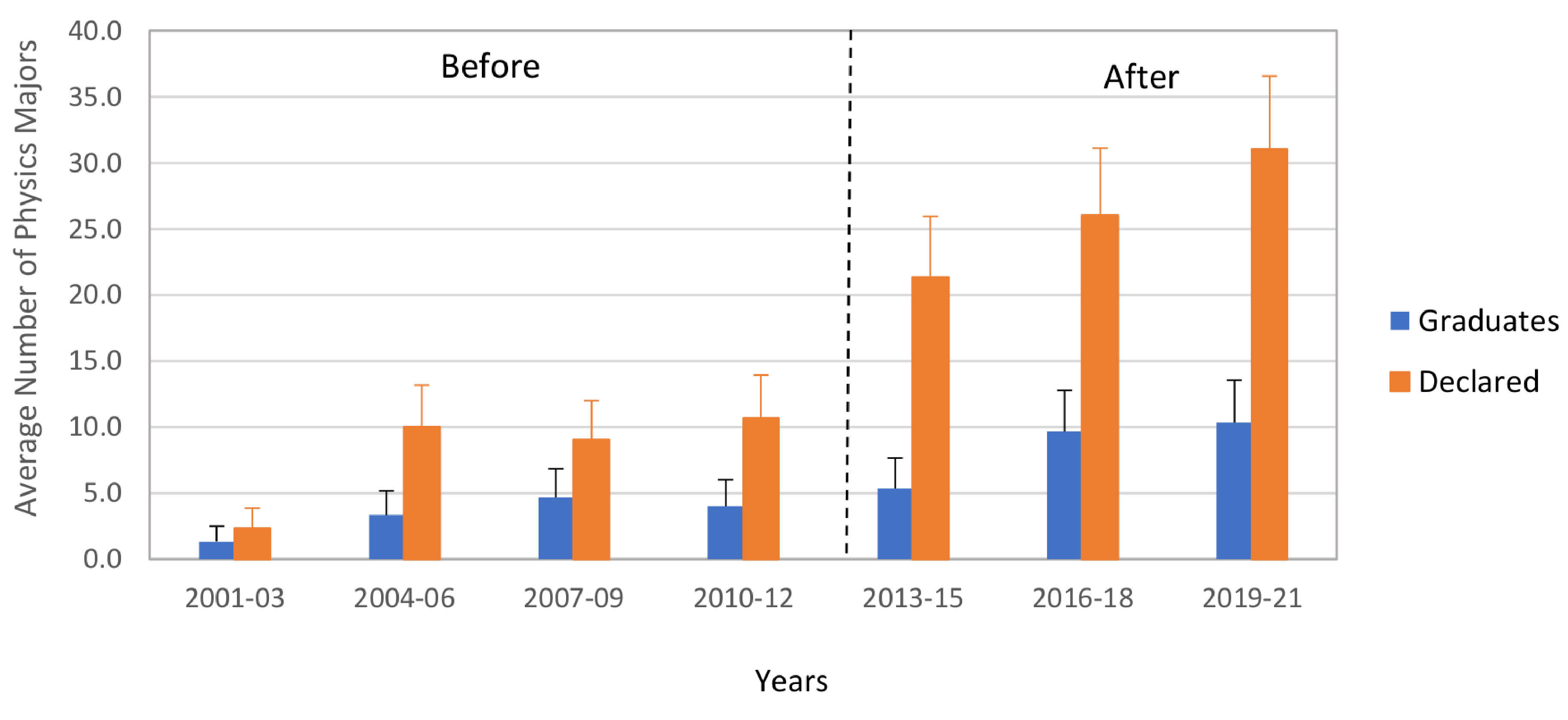}
\caption{Three-year averages for the numbers of declared physics majors (``declared'') and the number of undergraduate degree recipients (``graduates'').  Differences between the number of declared and the number of graduates are based on the definition of physics majors as discussed in the text (Sec.\ \ref{natlScape}). Vertical dashed line demarcates the beginning of the innovation initiation.}
\label{fig:Figure3}
\end{center}
  \end{figure}

Figure \ref{fig:Figure3} displays the three-year averages for the number of physics majors officially declared and bachelor's degree recipients. The data covers a twenty-year time period, which extends significantly before the implementations were added.  As a reminder, the data column ``2013--2015'' covers the initiation period for several innovative elements, including the first-year colloquium.  The following averages are more revealing when the innovative elements permeate a physics major's full matriculation.  Specifically, from 2016--2021, the average number of graduates was 9.8, while previously it was 3.9 (2001--15).  Not only did the number of graduates increase by more than two-fold, but the yearly fluctuation stabilized considerably. 

One immediate result of adopting an inclusive, cohort-building mindset was the admittance of Sophomores (second-year students) as declared majors.  This practice began in 2013 and helped explain the significant increase in ``declared'' column between 2012 and 2013.  Therefore, declared majors include second-, third-, and fourth-year students intentionally pursuing a physics degree. Prior to 2013, a `gate-keeping' mindset was more prevalent in the Physics Group, which required majors to show proficiency in upper-level coursework.  This change in mindset and practice afforded students earlier access as members of the cohort, including a greater sense of connection to the discipline.  This decision does not fully account for all the increases observed, since the number of graduates also increased significantly after 2016. 

Lastly, the noticeable increases of physics majors and graduates cannot be due to a weakening of the program or a loosening of accountability for its majors. During the time period covered in Figure \ref{fig:Figure3}, the number of credit units for the major remained roughly the same.  In fact, with the addition of the first-year colloquium and lab restructuring (c.f., Sec.\ \ref{curricular}), one could make the argument that the amount of coursework increased by at least one unit during the time period. Student expectations and engagement heightened due to programmatic augmentation, which was evidenced by recognition for both individuals (Goldwater Scholar) and the program (SPS Chapter Award).

\subsection{Increased Number of Women Degree Recipients}

\begin{figure}[!h]
  \begin{center}
\includegraphics[width=16.5cm,height=9cm]{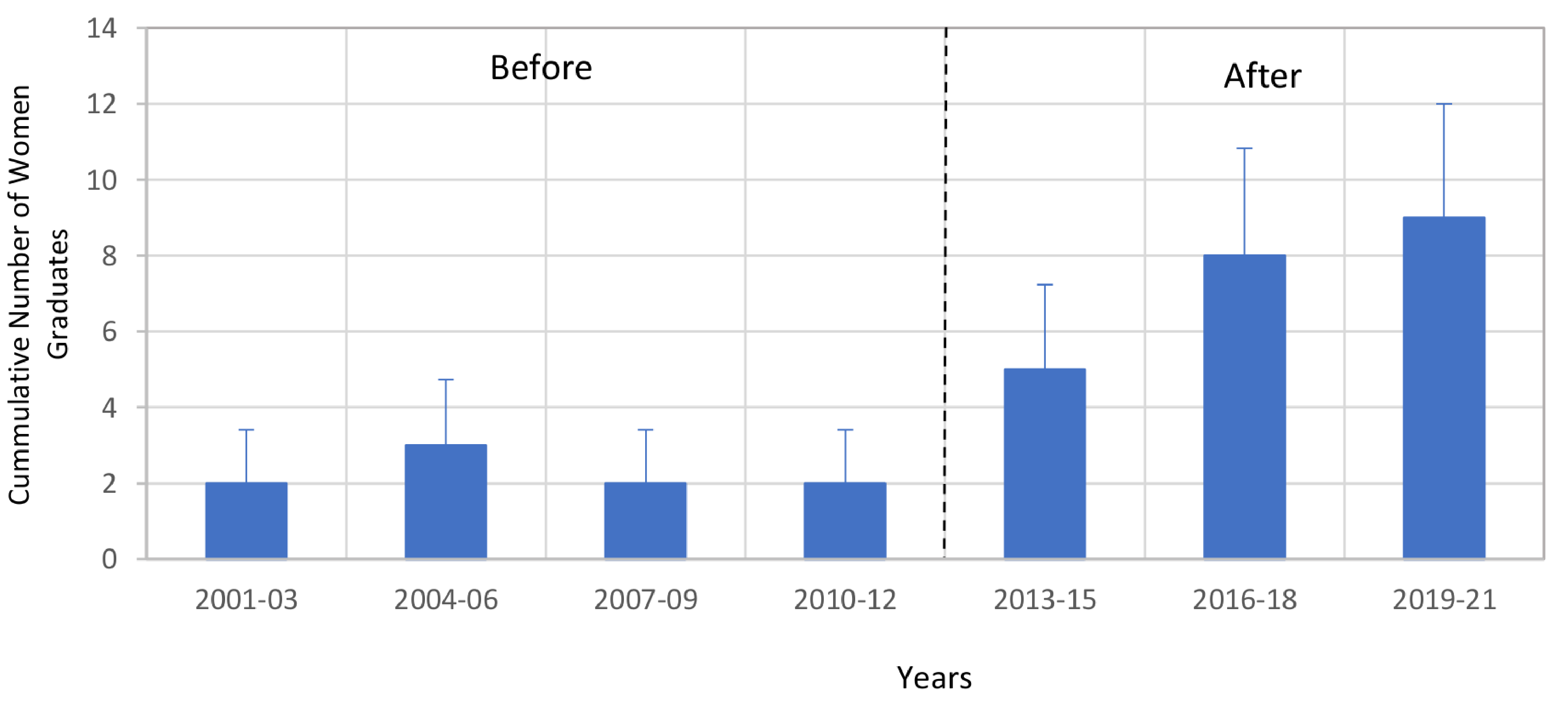}
\caption{Three-year, cumulative numbers of women who received an undergraduate physics degree in the Physics Group over a twenty-year period. Vertical dashed line demarcates the beginning of the innovation initiation.}
\label{fig:Figure4}
\end{center}
  \end{figure}

While the numbers of physics degree recipients have increased considerably over the last forty years, the percentages for women recipients remain approximately constant at 20$\pm5$\% since 2000 \citep{por19}.  Figure \ref{fig:Figure4} displays the number of women bachelor's recipients in the Physics Group over a twenty-year period.  Specifically, the yearly average of women graduating in physics from 2001-12 is 0.75, while the same average from 2013-21 is 2.4.  As a 200\% increase is weighed, a few considerations are addressed.

The undergraduate institution where the Physics Group is located enrolls a higher percentage of women undergraduate students, usually around 60\%, which is also the case for the period of the study.  However, prior to the study (2001-12), the average percentage of women was higher (closer to 65\%).  No particular alterations were made that would easily explain the significant rise in women physics graduates.  For the years of focused implementation (2013--21), the percentage of women graduates held steady around 30\%, which is about 10\% above the US average.  The retention of women in physics during the implementation period is now examined as a function of two innovative elements.  

\subsection{Increased Participation in Experiential Learning}

\begin{figure}[!h]
  \begin{center}
\includegraphics[width=13.5cm,height=9cm]{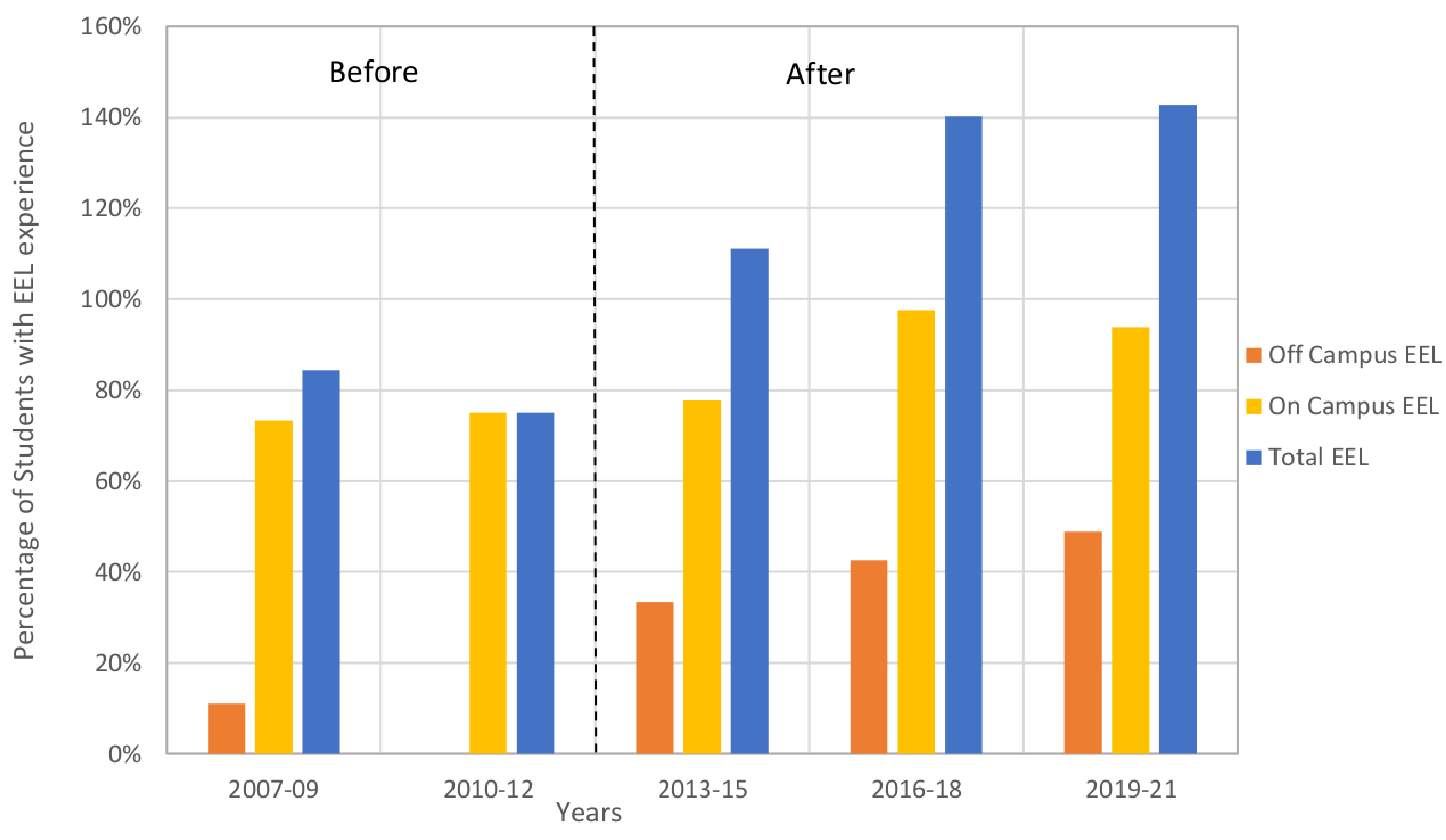}
\caption{Percent participation for all Physics Group majors in one (or more) EEL opportunities averaged over three-year periods.  Multiple experiences by the same student were tallied, therefore 100\% participation should not be interpreted as every major had a single experience. ``On Campus'' EEL include inter-disciplinary opportunities with programs like biology, chemistry, computer science, or mathematics.  ``Off Campus'' EEL include internships, REUs, and other collaborative partnerships.}
\label{fig:Figure2}
\end{center}
  \end{figure}

Over the last ten years, there has been a steady increase in the number of EEL opportunities from which our students have benefited. Figure \ref{fig:Figure2} displays data to support evidence of growth in the number of physics majors participating in one or more EELs. This includes all research performed by students on campus supervised by physics faculty, collaborative and/or inter-disciplinary research projects with other departments (chemistry, computer-science, mathematics), off campus research experiences such as NSF sponsored REUs and collaborative projects supervised by off campus mentors and internships. Several of our physics graduates often have more than one such EEL.   Between 2013-2018, the five-year average number of on-campus EEL for our students increased to 90\% compared to 69\% for the previous five years 2007-2011.  More starkly, the numbers of off-campus EEL for our majors has increased by a factor of six (7\% to 45\%) over a similar period.  Figure \ref{fig:Figure2} reflects the growth in each area over the time-period considered.  

\subsection{Retention of Women Physics Majors}
Prior to 2013, there was no consistent means of accounting for matriculation in the physics major as a function of original interest.  For example, there was no direct measure of incoming interest compared with the number of students who enrolled in the first semester of calculus-based physics (in Spring).  Therefore, the retention rates presented since the implementation period (2013) have no prior comparative data.  That said, it is still clear that the number of women retained within the physics major accord with the innovative elements introduced in the program (cf., Sec.\ \ref{innovateIntro}).

\subsubsection{With Respect to First-year Colloquium Enrollment}
\begin{figure}[!h]
  \begin{center}
\includegraphics[width=13.5cm,height=10cm]{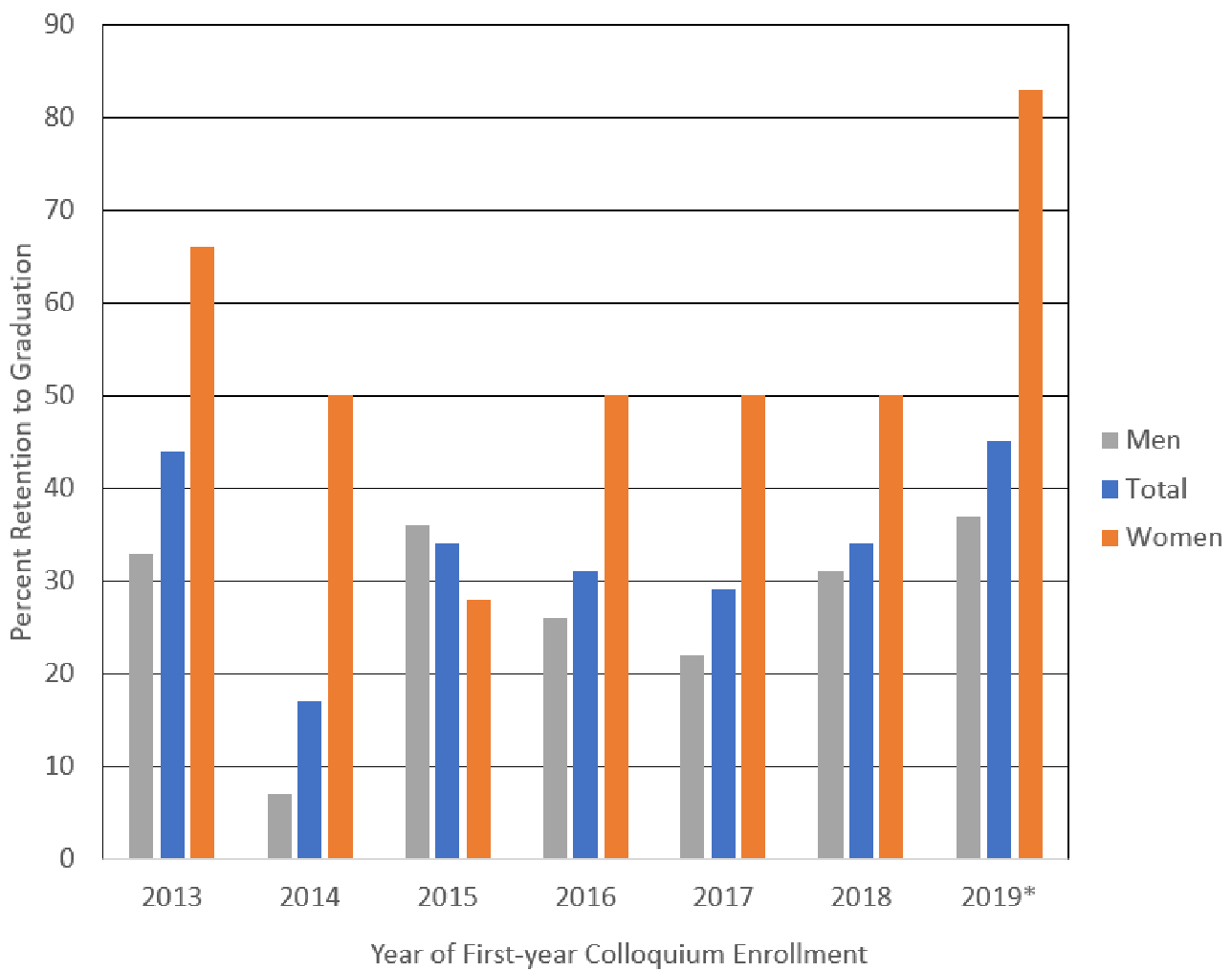}
\caption{Retention-to-graduation within the Physics Group for participants in the first-year colloquium by gender.  The percentages are based on final enrollments in colloquium and final numbers of degree recipients. The asterisk for 2019 indicates those who had not officially graduated before the completion of the study. Their retention was calculated after the third-year of completion.  }
\label{fig:Figure6}
\end{center}
  \end{figure}

Since 2013, a first-year colloquium was required for the physics degree, and it was strongly recommended by pre-admission advising to all incoming students who were interested in physics and/or engineering.  The enrollment for the course increased steadily for every pre-pandemic year, from 19 in 2013 to 36 in 2019.  Figure \ref{fig:Figure6} presents the retention percentages by gender for original enrollees who persisted until graduation.  Due to the frequency of course offering and higher enrollments, retention percentages are provided for each year during the study.  The asterisk for 2019 indicates those who had not officially graduated before the completion of the study, so their retention was measured prematurely (at the end of the third year).  Since it is in the third year that upper-level coursework begins, it is held with more certainty that a student would be retained. With the exception of a single year (2015), women were retained within the physics major at a higher percentage than men.  Therefore, during the time period of the study, women were retained in the physics major at a higher average percentage than men.

Even though the first-year colloquium continues, more recent numbers are not given here for two reasons. In Figure \ref{fig:Figure6}, 2019 is the last year that majors can confidently be reported as matriculating through the major, since a student's first- and second-year remain less certain. Second, since the course maintains a high relational component, pandemic effects are unclear for both institutional and colloquium enrollments.

\subsubsection{With Respect to Conference Participation}
\begin{figure}[!h]
  \begin{center}
\includegraphics[width=13.5cm,height=9cm]{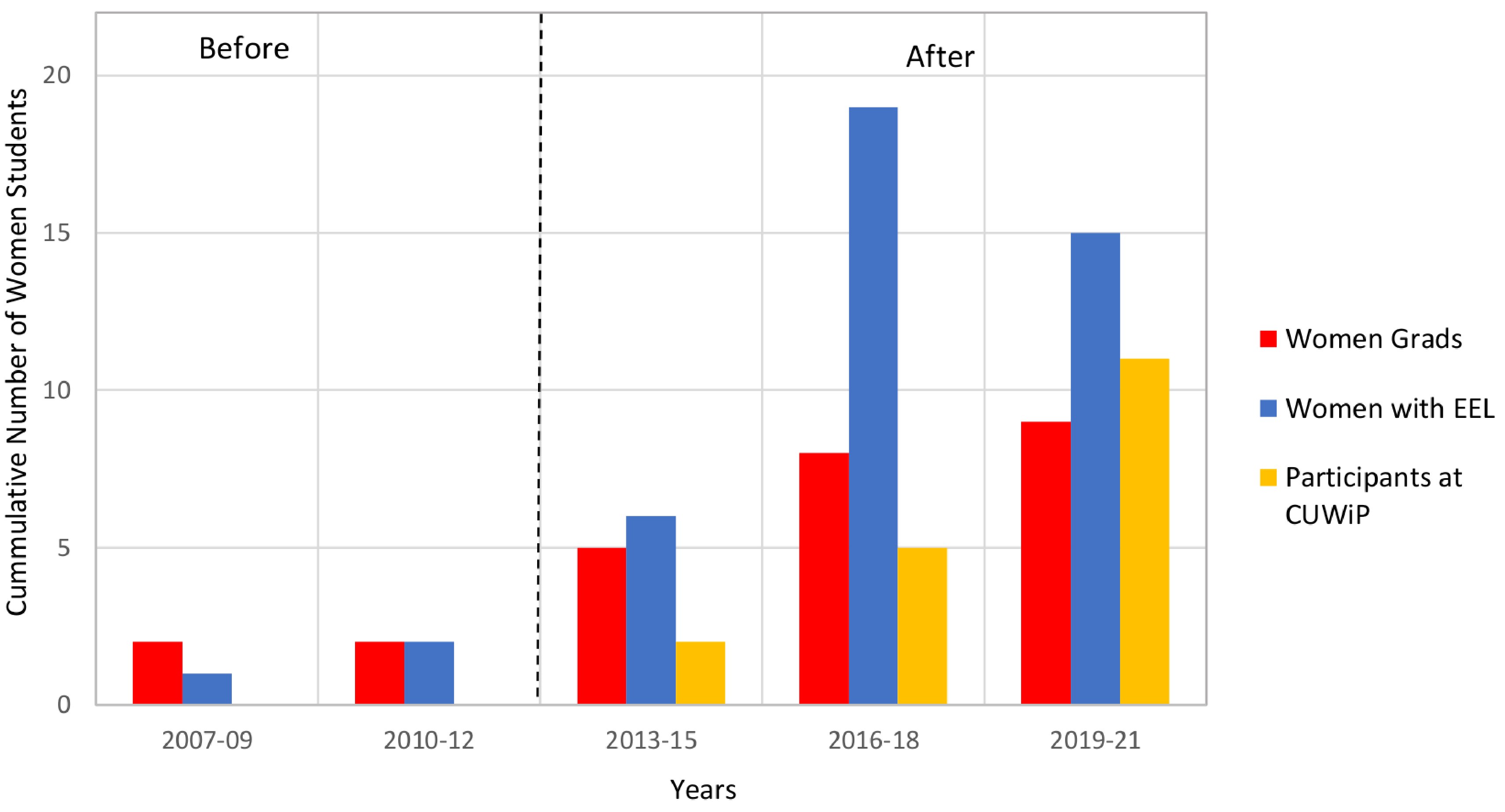}
\caption{Retention-to-Graduation for women in the Physics Group in comparison with EEL and CUWiP participation. Cumulative number of participants at the APS Conference for Undergraduate Women in Physics (CUWiP) by three-year grouping. Prior to 2013, there was no CUWiP attendance in the Physics Group.  If a participant attended more than once, it is only tallied at the first instance and suppressed for subsequent years. Multiple EEL for the same student were included in the tallies. }
 \label{fig:Figure1}
\end{center}
  \end{figure}

In particular, the Conference for Undergraduate Women in Physics (CUWiP) provided a timely and specific opportunity for women physics majors in the Physics Group.  These regional conferences are supported by the APS through funding from the National Science Foundation (NSF) and the Department of Energy (DOE). Typically, associated costs for conference participants are subsidized.  Figure \ref{fig:Figure1} shows the increase in numbers of women participants at CUWiP conferences, where each participant is counted only once.  These increases are independent and irrespective of the increases in total numbers of majors.

\subsubsection{With Respect to EEL Participation}

As a reminder, the following types of opportunities are included as EEL participation: REU experiences, on-campus research with STEM faculty, STEM-related educational experiences, and physics-related internships.  Students who participate in multiple EEL, particularly within historically underrepresented groups, are shown to positively correlate with self-evaluated STEM identity \citep{est18}.  Figure \ref{fig:Figure1} shows the relationship between women physics graduates and their participation in EEL opportunities.  The number of EEL accords with the increases in women physics degree recipients.  For Figure \ref{fig:Figure1}, multiple EEL for the same woman, physics major are included in the total, so only for the years ``2016--2018'' are there an average of $> 2$ EEL per woman graduate.  During the years of innovative element implementation (2013--21), there was greater than one EEL opportunity per woman graduate. For years prior to 2007, there was no EEL information cataloged.

\subsection{Retention of Women within STEM Careers}\label{womenSTEM}
As displayed in Figure \ref{fig:Figure4}, noticeable and consistent increases in women degree recipients begin during the ``2013-2015'' segment, which coincides with the initiation of most innovative elements.  This same three-year segment also marks the first time where the average number of physics majors eclipses 20.0, and total graduate average above 5.0 (c.f.\ Fig.\ \ref{fig:Figure3}.  Since immediately after graduation presents another pressure point for attrition from STEM fields, we maintain an interest in what these retained physics graduates do after they receive their degree. 

\begin{figure}[!h]
  \begin{center}
\includegraphics[width=13.5cm,height=7.5cm]{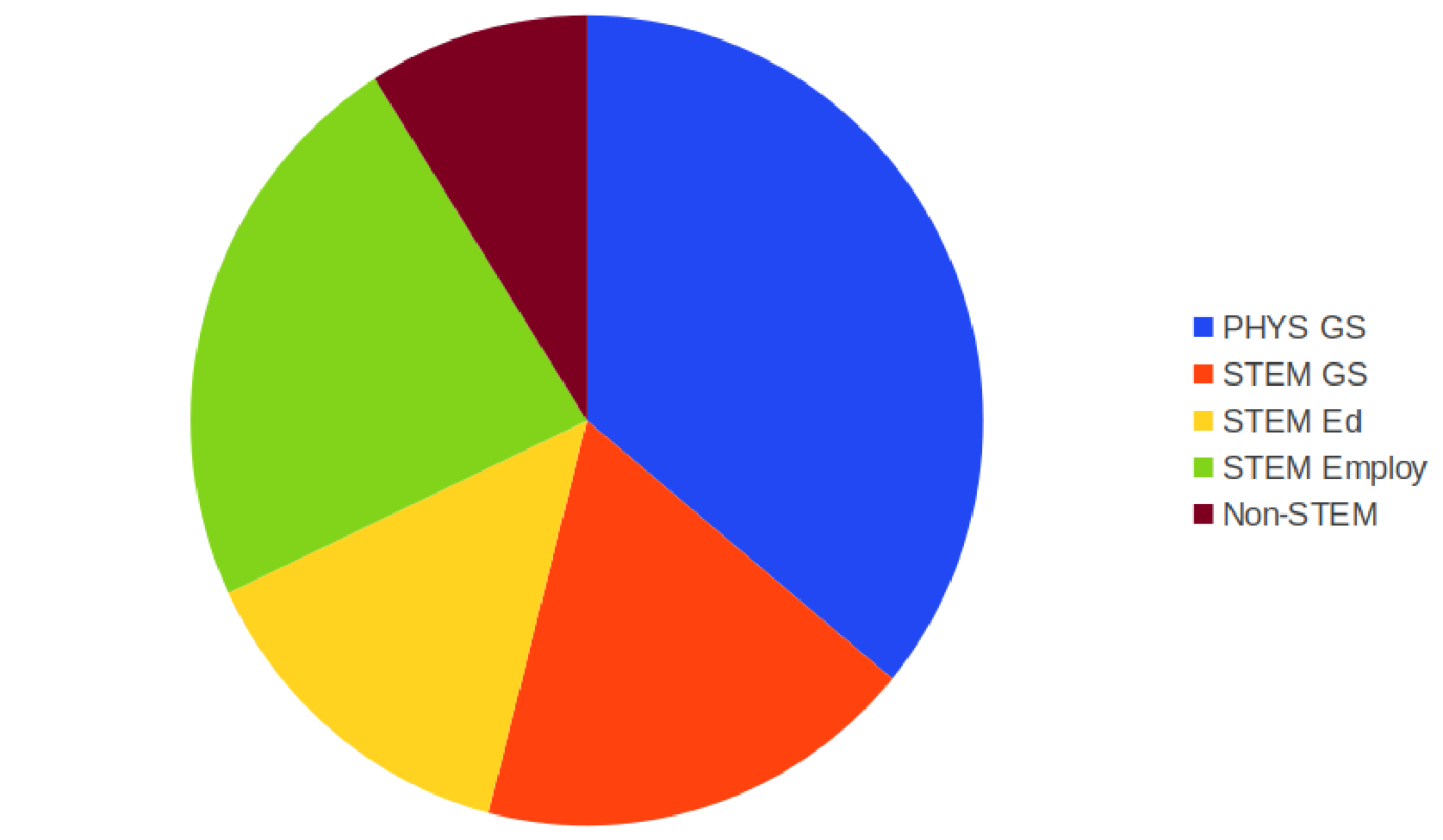}
\caption{Post-graduate career trajectory decisions for twenty-two women degree recipients from the Physics Group in the years 2013--2021. Decisions were categorized one-year after graduation.  All but two of the graduates were retained within STEM advancement positions. Legend categories are discussed in the text (Sec.\ \ref{womenSTEM}).}
\label{fig:Figure8}
\end{center}
  \end{figure}

Figure \ref{fig:Figure8} presents the decisions for twenty-two women degree recipients from 2013--21, one-year after graduation.  Decisions are separated loosely along the following categories: Physics-related graduate school (PHYS GS), non-Physics, STEM-related graduate school (STEM GS), STEM-related employment professions (STEM Employ), STEM-related education professions (STEM Ed), and non-STEM employment (non-STEM).  Of the twenty-two women, less than 10\% (2/22) persist in a career not directly related to STEM advancement.  Here, advancement is defined as involvement in STEM research and support (STEM Employ), learning (GS), or education (STEM Ed).  Therefore, women graduates from the Physics Group matriculate into STEM-related trajectories at $\sim$90\% during the 2013--21 time period.

Graduate training, whether physics-related or STEM broadly, comprised over half of the post-graduate decisions (12/22) for these women.  Physics-related training (PHYS GS) included physics subdisciplines, engineering, and materials science.  More broadly, STEM-related, graduate training (STEM GS) included fields like computer science, applied math, and veterinary medicine. A few of the women graduates (3) completed their graduate training since leaving the Physics Group, and all three matriculated into STEM-related employment. Moreover, at the time of publishing, all of these women graduate students remained in their programs, ranging from 1--6 years later, or graduated from them.

When considering the entire group of women bachelor recipients from 2013-2021, a more broad statement emerges about their longevity within STEM advancement careers. Through continuing to follow the post-graduate trajectories of the twenty women already involved in STEM advancement, we find that all of them continue to find meaningful, STEM-related employment, even up to eight years after graduation, beginning in 2013. Even though some of them have transitioned into parenthood, parallel fields, and/or promotion, these women continue to thrive within STEM-advancement careers. 

\section{Discussion}
To summarize, several innovative elements augmented the undergraduate program in the Physics Group over a period of eight years. With a staggered initiation, these additions were categorized as curricular, co-curricular, or experiential (Sec.\ \ref{innovateIntro}).  Some of these elements were introduced as a result of national organization documentation, some from STEM literature, and some were novel within the Physics Group.  Changes and growth in the program were noted over the same eight-year period, from 2013--2021. There were three primary results that accorded with the addition of these innovative elements. First, the total number of majors and graduates increased by approximately 200\% compared with the previous thirteen years (Fig.\ \ref{fig:Figure3}).  Second, the number of women in the major were retained at a higher rate than men (Fig.\ \ref{fig:Figure6}). Third, these women graduates were employed in STEM-advancement positions at $\sim$90\%, from one-year after graduation and extending out to eight years post-graduation (Fig.\ \ref{fig:Figure8}).  Since the implementation and evaluation period of the elements overlapped, it was impossible to determine which elements contributed specifically to each result.  That said, we provide a guiding analogy to frame an interpretative discussion of these results within the context of programmatic change.

\subsection{Leaky Pipelines and Delta Distributaries}
Historically, ``leaky pipelines'' refer to losses in STEM representation, ranging from  aggregate \citep{nifa14} to minoritized populations \citep{liu19}, and even both \citep{met10}.  More specifically, the analogy pertains to a decrease in STEM participation for women, initially beginning in transitions from secondary to undergraduate \citep{arc16}, then extending to PhD representation \citep{mil15} and careers in academia \citep{she14}. Systemic problems are prevalent with historically underrepresented groups that are more clearly recognized at formal transitions, though other alternatives are suggested \citep{rai18,wit20}. Logically, since the numbers of men and women in K-12 are roughly equal, including the numbers of students taking physical science coursework in high school, then there should be roughly equal numbers of women and men in STEM careers.  For women in physics and engineering, there are greatly reduced numbers of bachelor's recipients, which is interpreted as a ``leak.''  Sometimes the cause of a leak is focused on gender discrimination \citep{gro18} and sometimes more broadly on the nature of science \citep{bli05}. However, questions about the analogy have been raised, either as to its efficacy toward improvements \citep{can14} or its over-simplicity \citep{hin20}.

The innovative changes implemented in the Physics Group offer a more inclusive and branched approach than a pipeline analogy. By cultivating a `hidden physicist’ model, the program encourages students to view physics as a pathway to diverse STEM careers. Perhaps, a more flexible image is more helpful. Specifically, the Physics Group considers a river delta analogy where distributaries (post-graduation decisions) are initially kept broad within a primary tributary (major program) but allowed to spread and disseminate at the delta by the third- and fourth-years. Figure \label{fig:Figure9} shows a schematic of a typical river delta. While not completely accurate in every detail, we note that river delta plains become fertile areas for new growth. By leaving unanswered the question, ``What can a physics major do?,'' any number of post-graduation options are encouraged \citep{fle18}.  We examine a river delta analogy within two important concepts, STEM identity and programmatic structure.     

\begin{figure}[!h]
  \begin{center}
\includegraphics[width=8.5cm,height=7.5cm]{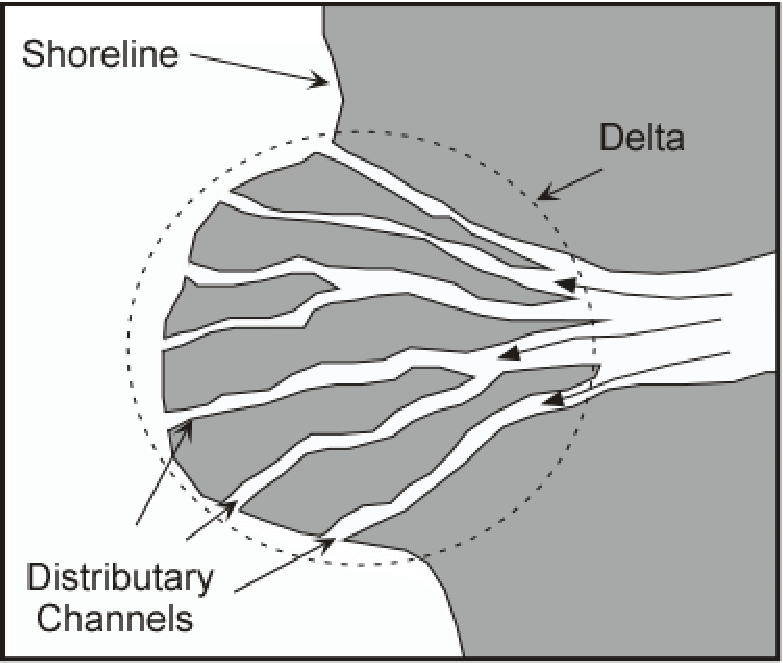}
\caption{A typical schematic of a river delta.  The primary tributary (right side) branches into several distributaries as it nears the much larger body of water.  The Physics Group image of a typical undergraduate's matriculation through the program loosely follows this image.(\tiny{\url{https://www2.tulane.edu/~sanelson/Natural\_Disasters/riversystems.htm}}) }
\label{fig:Figure9}
\end{center}
  \end{figure}

\subsection{Programmatic Structure, Building the Banks of the Tributary}
Within a river delta analogy, the number and strength of the distributaries depends on the tributary.  In our analogy, the augmented physics program serves as the tributary and includes student majors with faculty.  The innovative elements discussed in Section \ref{innovateIntro} provide fortified banks for the increased number of majors.  To be clear, `fortifying' does not mean constraining individual choices, whereby students may choose to leave the program and/or switch majors.  In the establishment of STEM identity, freedom and encouragement must be given as students persist through difficult circumstances \citep{fle21}. By drawing on best-practices from PER literature, in combination with unique innovations specific to the Physics Group \citep{hen15}, a well-fortified tributary is kept. Two implemented examples from the physics program confirm the imagery of building strong banks for women in undergraduate trajectories.

The first-year colloquium serves as the tributary entry point.  Keeping the course in an exploratory state, while also introducing a broad diversity of those who are in the Physics Group, seems to keep undecided participants in an interested state.  The connection seems clear between the increase of majors and graduates overall (Fig.\ \ref{fig:Figure3}) and the introduction of the first-year colloquium (Fig.\ \ref{fig:Figure6}).  An open award structure for physics excellence rewards the unique contributions and accomplishments of many who persevere in one of the most rigorous disciplines (Sec.\ \ref{innovateIntro}), while early participation within many informal programs sponsored by the Physics Group helps to establish identity \citep{ret21}.  By continuing to keep the tributary entry broad, a student comes to the Junior Review event with a greater opportunity of successful STEM experiences.

Synergizing CURE results from the literature with the curriculum structure in the Physics Group was an innovation beyond documentation \citep{cor15,woo18,reic19}. With the implementation of a CURE-based laboratory, students were able to receive some partial instruction in all upper-level electives, but it also facilitated further EEL persistence.  By implementing a half-credit research/independent study experience that usually precedes a larger, full-credit summer experience, the Physics Group better connects research experiences between the coursework and beyond.  These pre-experiences parlay well and create inertia for a student's formative summer EEL opportunities (particularly after their third-year). Such opportunities allow scaffolding of skills and strengthening a knowledge-base through a research experience, which promote a heightened persistence in undergraduates \citep{est11}. All of these EEL experiences serve to fortify the banks of the program tributary facilitating students to successful STEM trajectories post-graduation. 

\subsection{STEM Identity, Flooding the Delta}
Many of the innovative elements reinforce the development of interpersonal and intrapersonal factors known to affect retention in the STEM disciplines, such as science (STEM) identity and sense of belonging. As science identity is predictive of longer-term persistence in STEM-related fields \citep{ere21}, and sense of non-belonging is cited as a reason for “leaving” the sciences (especially for women, \citealp{lew17}), these aims are essential to the presented study and well-supported in the literature. Specifically, \citet{est18} note that completing multiple semesters of research and/or internship within STEM correlates strongly with establishing identity within STEM. Related, \citet{fin17} demonstrate that participation in an intensive week-long co-curricular program just as students enter college, including engaging with peers with similar interests in addition to several elements related to reflecting on goals and career aims and getting connected to faculty, is related to increased science identity and sense of belonging. In addition, increases in science identity seem to be driven by increases in belonging (to STEM and institutional community; \citealp{kuc19}), suggesting that efforts such as providing greater connection to faculty and peers from early on in the program facilitate such development.   As research suggests, formation of identity is iterative as students grow in confidence through repeated interaction with content, problem-solving, and experimental techniques \citep{kea18}.  It would be unsurprising if women were better established as shareholders and valued members, where the numbers of women graduates increased.

To substantiate a distributaries analogy within the Physics Group, measures in Figure \ref{fig:Figure6} seem to reduce attrition at a high school to undergraduate transition, since women are retained in the major at higher rates than men. Specifically, if a high school student shows high aptitude and interest in physics (so-called, ``Exceptional Physics Girls'', \citealp{arc16}), then she is retained at rates above national averages in the Physics Group.  The result in Figure \ref{fig:Figure6} accord with an attempt to keep the primary tributary as broad as possible, early within the undergraduate experience.

Similarly, if a woman receives her bachelor's degree in the Physics Group, then her STEM identity seems more solidified as she moves into a STEM advancement position (Fig.\ \ref{fig:Figure8}).  When examining the approximate percentages shown in Figure \ref{fig:Figure8}, it is clear that a greater number of women in STEM advancement occupy the non-PHYS GS categories.  That is, fewer women are retained to become (traditional) physicists than not.  However, the Physics Group still considers this successful, since the goal is to flood the delta.  A `flooded delta' represents scenarios where there is a greater likelihood that young women (K-12 students) will see someone like them in STEM-advancement positions. Without becoming elitist, the Physics Group believes that `hidden physicists' critically participate in other non-physics, STEM fields, because physics provides a unique way of knowing \citep{mar13}.   These unique pathways serve to diversify the face of science by increasing the number of distributaries for physics bachelor recipients.

\section*{Conflict of Interest Statement}
The authors declare that the research was conducted in the absence of any commercial or financial relationships that could be construed as a potential conflict of interest.

\section*{Author Contributions}

RB and MCF contributed to conception and implementation of programmatic innovation. MCF organized and collected the pertinent data. RB and MCF constructed the histograms. MCF wrote most of the manuscript's first draft, where DFVN provided significant contribution. DFVN contributed to the construction of STEM identity and belonging framework. All authors contributed to manuscript revision, read, and approved the submitted version.

\section*{Funding}
MCF thanks the Roanoke College sabbatical program for its generosity in reassign-time.  MCF also thanks the Margaret Duke Endowment at the University of Mary Washington from which reassign-time was provided.

\section*{Acknowledgments}
All Authors appreciate the Office of Institutional Research and the Registrar at Roanoke College for their gracious cooperation.

\bibliographystyle{Frontiers-Harvard} %  Many Frontiers journals use the Harvard referencing system (Author-date), to find the style and resources for the journal you are submitting to: https://zendesk.frontiersin.org/hc/en-us/articles/360017860337-Frontiers-Reference-Styles-by-Journal. For Humanities and Social Sciences articles please include page numbers in the in-text citations 
\bibliography{mcf2}

%%% Make sure to upload the bib file along with the tex file and PDF
%%% Please see the test.bib file for some examples of references

\end{document}